\documentstyle[twocolumn,psfig,prl,aps]{revtex}

\begin{document}

\draft
\twocolumn[\hsize\textwidth\columnwidth\hsize 
\csname @twocolumnfalse\endcsname

\title{Asymptotically exact solutions of Harper equation}

\author{A. G. Abanov, J. C. Talstra}
\address{James Franck Institute
of the University of Chicago,
5640 South Ellis Avenue, Chicago, Illinois 60637}

\author{P. B. Wiegmann}
\address{James Franck Institute
and Enrico Fermi Institute
of the University of Chicago,
5640 South Ellis Avenue, \\
Chicago, Illinois 60637 \\
and
Landau Institute for Theoretical Physics}

\maketitle

\begin{abstract}
We present asymptotically exact solutions of an incommensurate Harper
equation---one-dimensional Schr\"odinger equation of one particle on a
lattice in a cosine potential. The wave functions can be written as an
infinite product of {\em string polynomials}. The roots of these
polynomials are solutions of Bethe equations. They are classified
according to the {\it string hypothesis}. The string hypothesis gives
asymptotically exact values of roots and reveals the hierarchical
structure of the spectrum of the Harper equation.
\end{abstract}

 ]

1. Incommensurate quantum systems present a broad variety of complicated
fractal objects and strange sets. The Harper
equation (or almost Mathieu equation) is one of the most notable:
\begin{equation}
 \label{harper}
    \psi_{n+1}+\psi_{n-1}
    +2\lambda\cos(\theta+2\pi n\eta)\psi_n=E\psi_n 
\end{equation}
It has two competing periods $\eta^{-1}$ and 1, and describes a particle
on a one-dimensional lattice in a periodic potential. The same equation
(\ref{harper}) also describes a particle on a two-dimensional square
lattice in a uniform magnetic field $\eta$ flux quanta per elementary
plaquette (Azbel-Hofstadter problem). Then $\lambda$ is an anisotropy of
the square lattice and parameter $\theta$ is a Bloch multiplier.

When $\eta$ is an irrational number, i.e., the period of the potential is
incommensurate with the period of the lattice, the spectrum and wave
functions are known to have a rich hierarchical
structure\cite{Azbel64}-\cite{LastJit94}. The physics of the problem
depends on the value of parameter $\lambda$ and falls into three
universality classes. For $\lambda<1$ all wave functions are extended
and the energy spectrum has a non-zero Lebesgue measure
$4(1-\lambda)$. For $\lambda>1$ all wave functions are
localized (in this letter we consider only ``typical'' irrational numbers
which can not be approximated by rationals too well)
and the energy spectrum has zero measure. At the most interesting
``critical'' point $\lambda=1$ wave functions neither localized nor
extended, but exhibit a power law scaling, while the spectrum is
singular continuous (a set without isolated points but with zero
Lebesgue measure).

Since the empirical observations of Hofstadter \cite{Hofstadter76}, the
evidence has been mounting that these spectra are regular and universal
rather than erratic or ``chaotic''. Few years ago it has been
shown\cite{WiegmannZab94,FadKash95,Kutz94} that despite the complexity of
the spectrum, the Harper equation at any rational $\eta=P/Q$ is
integrable and can be ``solved'' by means of  the Bethe Ansatz (BA).
This had opened the possibility of describing the complex behavior of an
incommensurate system as a limit of a sequence of integrable models. In
this letter we analyze the Bethe equations\cite{HKW94} for $\lambda=1$
and show how the (BA) reveals the hierarchical structure of the spectrum.
We show that 
(i) solutions of the BA equations ({\it roots}) fall into
complexes---{\it strings}; 
(ii) the number of roots in a string ({\it length of the string}), known
as Takahashi-Suzuki numbers\cite{TakSuz72} are intimately related to 
the Hall conductance of a state; 
(iii) strings provide an asymptotically exact wave functions for the
entire spectrum.

To introduce the scaling concept let us
consider an irrational $\eta$ as the limit of rational approximants 
$\eta_j = \frac{P_j}{Q_j}$. With rational number $\eta_j$ instead of
$\eta$ the potential in (\ref{harper}) becomes commensurate with the
lattice and the problem is reduced to diagonalization of a $Q_j\times
Q_j$ matrix. The spectrum consists of $Q_j$ bands separated by $Q_j-1$
gaps\cite{Mouche89}. In this letter we show that the wave function of the
problem with rational $\eta_j$ can be written approximately as a
product of a wave function for the $\eta_{j-1}$ (ancestor wave function)
and some polynomial which roots form a so-called {\em string}. By
recursion the wave function with irrational $\eta$ appears as an
infinite product of {\em string polynomials}. The multiplicative
structure thus obtained is asymptotically exact (see
eq.\ref{PsiAnsatz}). It resembles the multiplicative formulae
conjectured in\cite{ThoulessNiu83}.

2. {\it The Bethe Ansatz.}
If $\eta=P/Q$ is a rational number then the solution of  (\ref{harper})
can be chosen as Bloch function $\psi_n=e^{ik_x n}\chi_n$ where
$\chi_n=\chi_{n+Q}$. Although the BA solution is known for arbitrary
values of Bloch multipliers $k_x$ and $k_y\equiv\theta$\cite{FadKash95},
below we consider only the points of Brillouin zone $(k_x,k_y)$ which
correspond to edges and centers of bands. At these ({\em rational})
points the BA solution is particularly simple\cite{comment}.  BA
solution can be obtained in two steps. The first step is the isospectral
transformation:
\begin{equation}
 \label{psi}
    \psi_n=\sum_{m=0}^{M} c_{nm}\Psi(-i\rho q^m) 
\end{equation}
to the difference equation:
\begin{eqnarray}
    \mu\kappa z E{\Psi}(z)
    &=& iq (z +i\tau\kappa q^{-\frac{1}{2}}) 
    (z -i\kappa q^{-\frac{1}{2}}) {\Psi}(qz) 
 \nonumber \\
    &-& iq^{-1} (z -i\tau\kappa q^{\frac{1}{2}}) 
    (z +i\kappa q^{\frac{1}{2}}) {\Psi}(zq^{-1}),
 \label{HamChiral223}
\end{eqnarray}
where $\rho=\exp(i\frac{k_x+k_y-\pi P}{2})$, $M=2Q-1$ if $P$ is odd and
$M=Q-1$ if $P$ is even; $q=e^{i\pi\eta}$. The choice of $\tau=\pm 1$ 
yields either levels at the center or at the edges of bands
respectively. The rational points are degenerate and are labeled by
discrete parameters $\kappa,\mu=\pm 1$. The coefficients in (\ref{psi})
are  given by ``quantum dilogarithms''
\begin{equation}%
    c_{nm}=\prod_{j=0}^{m-1}\Big(e^{ik_y}q^{2n+1/2}
     \frac{1+i\tau\kappa\rho^{-1} 
    q^{-j-1/2}}{1-i\kappa\rho q^{j+1/2}}\Big). 
\end{equation}%

The second step is based on the  integrability of
eq.(\ref{HamChiral223}): $Q$ its (regular) solutions  are polynomials
of $z$ of the degree $Q-1$:
\begin{equation}
 \label{polform}
    \Psi(z)=\prod_{i=0}^{Q-1} (z-z_i).
\end{equation}
The roots $\{z_i\}$ satisfy the system of Bethe equations: 
\begin{equation}
    q^{Q}\prod_{k=1}^{Q-1}\frac{qz_i-z_k}{z_i-qz_k} 
    = \frac{\left(z_i-i\tau
    \kappa q^{\frac{1}{2}}\right)
    \left(z_i +i\kappa q^{\frac{1}{2}}\right)} 
    {\left(q^{\frac{1}{2}}z_i
    +i\tau\kappa\right) \left(q^{\frac{1}{2}}z_i -i\kappa\right)}.
 \label{BetheAnsatz}
\end{equation}
Solutions of the BA equations give the wave functions of the
Harper equation at band's edges and centers. Their energy is given by
\begin{equation}
 \label{en1}
     E=i\mu q^{Q}(q-q^{-1})
    \left[\kappa\sum_{i=1}^{Q-1}z_{i} 
    -i\frac{1 -\tau}{q^{1/2}-q^{-1/2}} \right]. 
\end{equation}
At first glance, the BA equations
(\ref{BetheAnsatz}) look even more complicated than the original Harper
equation. However, the BA
equations (\ref{BetheAnsatz}) provide a better description of the problem
in the most interesting, incommensurate, limit $Q\rightarrow \infty$.
In a sense, the incommensurate limit plays the role of the thermodynamic
limit, with $Q$ being the size of the system.

3. {\it String hypothesis}.
Below we formulate the string hypothesis which allows us to obtain the
approximate solutions of the BA equations. The hypothesis is based on
the analysis of singularities of the BA, and is supported by extensive
numerics. We present a detailed analysis of singularities in a more
extended publication\cite{ATW97}. Here we just formulate the string
hypothesis and present some immediate consequences. First, we
need the notion of {\it strings}, a {\it Hall conductance}, a
{\it hierarchical spectral flow}  and a {\it hierarchical tree}.

a) A string of spin $l$, parity $v=\pm 1$ and center $x$ is a set
of $2l+1$ complex numbers:
\begin{equation}
 \label{stringdef}
     vx \{ q_l^{-\frac{l}{2}},
    q_l^{-\frac{l}{2}+1},\ldots, q_l^{+\frac{l}{2}} \},
\end{equation}
where $x$ is a real positive number and $q_l^{2l+1}=\pm 1$.

b) The Hall conductance $\sigma(k)$ of the band $k=1,\ldots, Q$
is the Chern class of the band, i.e., the number of zeros of the wave
function as a function of $k_x,k_y$ in the Brillouin zone. We will also
need the Hall conductance of $k$ filled bands: $\sigma_k$. The latter is
assigned to the $k$-th gap, so that $\sigma(k)=\sigma_k-\sigma_{k-1}$.
They are determined by the Diophantine equation\cite{TKNN82,DAZ85}
\begin{equation}
 \label{DE}
    P \sigma_k = k\; (\mbox{mod}\; Q)
\end{equation}
restricted to the range $-Q/2<\sigma_k\leq Q/2$.

c) {\it Hierarchical spectral flow}.
We refer to the band $k'$ of the problem with parameter
$\eta'=P'/Q'$, as to the parent band of the parent generation
corresponding to the daughter band $k$ of the daughter generation
$\eta=P/Q$ if: 
(i) the Hall conductance of the band $k$ is the ratio
between the difference of the number of states of parent and
daughter bands and the difference between the periods of generations
\begin{equation}
 \label{integrstreda}%
    \frac{1}{Q}-\frac{1}{Q'}=\sigma(k) (\frac{P}{Q}-\frac{P'}{Q'}),
\end{equation}%
(ii) $\eta'$ and $\eta$ are closest rationals with $Q'\le Q$, and
$(k'-1)/Q'<(k-1/2)/Q<k'/Q'$. These conditions  may be viewed as the
integrated Streda formula (see Ref. \cite{Streda82}), which determines
the spectral flow while varying the period from $\eta'$ to $\eta$.
Starting from some particular band $k$ of the generation $\eta$ one can
obtain from Eq.(\ref{integrstreda}) the sequence of ``ancestor'' bands
belonging to generations $\eta_j\equiv \frac{P_j}{Q_j}$---subsequent
rational approximants to $\eta$. The bands belonging to different
periods $\eta_j$ together with parent-daughter relations the
{\em hierarchical tree}. Let us notice that the sequence $\eta_j$ differs
from the  approximants obtained by truncating the continued fraction
expansion of $\eta$. Instead, it consists of {\it intermediate
fractions}\cite{Venkov70}. Let us also notice, that parents of states of
a given generation do not necessarily belong to the {\em same} 
generation.

Now we are ready to formulate the {\it string hypothesis}. At large $Q$
each solution of the BA consists of strings, so that each state can be
labeled by spins $\{l_j,l_{j-1},\ldots\}$ and parities
$\{v_{j},v_{j-1}, \ldots\}$ of strings, such that the total number of
roots $\sum_{i=1}^k (2l_i+1)=Q-1$. We refer to the set of lengths and
parities of strings constituting the solution for a given energy level
as to a string content of this level. The length of the longest string
in a string content of a given energy level is the Hall conductance of
the corresponding energy band: $2l+1=|\sigma (k)|$. The ``period'' of
this string $q_l=e^{i\pi\eta_l}$ is uniquely determined by the
requirement that $\eta_l=\frac{P_l}{2l+1}$ is the best approximant for
the period $\eta$ of the state. The parity of the longest string is
$v_l = -iq_l^{l+1/2}\kappa=(-1)^{\left [\eta l\right]}\kappa$.
The remaining roots are given by the solution of the BA equation for the
parent state of the parent generation. In other words
\begin{equation}
    \Psi^{\rm daugther}(z)\approx\Psi^{\rm parent}(z)
    \prod_{m=-l}^{l}(z-x_lv_l
    q_{l}^m).
 \label{PsiAnsatz}
\end{equation}
The accuracy of eq.(\ref{PsiAnsatz}) is determined by the accuracy of
approximation of $\eta$ by $\eta'$ and $\eta_l$ which is of the order of
$1/l^2$ ($\eta$, $\eta'$ and $eta_l$ are all rational approximants of
some ``typical'' irrational number). Thus the formula (\ref{PsiAnsatz})
is asymptotically exact. It gives the roots of the largest string of the
length $2l+1$ with the accuracy $1/l^2$ which gets higher along the
hierarchical tree when $\eta\rightarrow\;{\rm irrational}$.

The complete string composition of a
given state can be obtained by means of the iterative procedure.
Starting from a state of the generation
$\eta$, first we find the branch of the hierarchical tree that this state
belong to. Then we determine the Hall conductances of all states of the
branch down to the origin. This determines the lengths, periods and
parities of strings.

The strings hierarchy has been  obtained through the analysis of
singularities of BA (see\cite{ATW97}). As it was expected a set of
possible lengths of strings is a set of Takahashi-Suzuki numbers
(dimensions of irreducible representations of $U_q(SL_2)$ with definite
parity\cite{MezNep90}). Eq.(\ref{DE}) provides a relation between them
and Hall conductances.

The iterative procedure provides an algorithm for writing down the wave
functions of all states for any rational $\eta$. The only unknowns are
the centers of strings. They, however approach 1 with accuracy ${\cal
O}(1/l)$. The accuracy of the recursive eq.(\ref{PsiAnsatz}) is
${\cal O}(l^{-2})$. The string content of a state (i.e., lengths and
parities of strings) is a topological characteristics, while the
centers of strings are not.

\begin{figure}
\centerline{\psfig{file=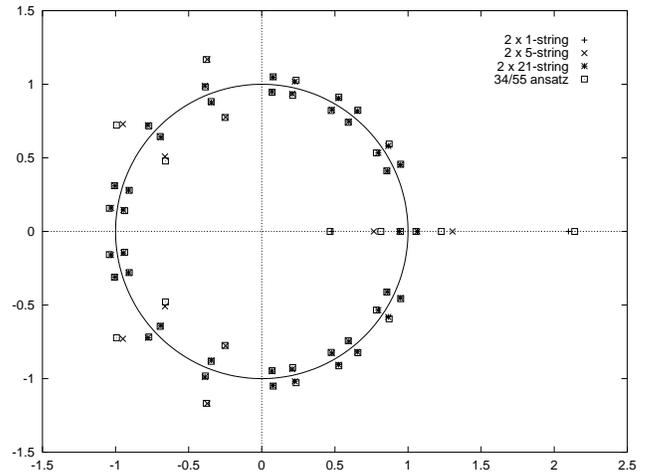,width=3.5in,angle=-90}}
\vspace{0.5cm}
\caption{Shown in the figure are the roots of  the lowest rational state
of the Harper equation at $\eta=34/55$. The roots are organized in
strings: two of the length 1($+$), two of the length 5($\times$) and two
of the length 21($*$). These roots can be approximated by ansatz
(\protect\ref{emp-wf-bottom}) and are plotted in the Figure as
boxes(\fbox{$\cdot$}). Roots tend to repel each other. This ``splitting'' is of
the order of ${\cal O}(1/l)$.
}
\label{fig.3455}
\end{figure}

To illustrate the iterative procedure let us consider the bottom edges of
the lowest band of the spectrum and choose $\eta=\frac{\sqrt{5}-1}{2}$ to
be the golden mean. The sequence of rational approximants is given by
ratios of subsequent Fibonacci numbers $\eta_i= \frac{F_{i-1}}{F_i}$,
where the $F_i$ are Fibonacci numbers ($F_i=F_{i-2}+F_{i-1}$ and
$F_0=F_1=1$). The set of Hall conductances = Takahashi-Suzuki numbers =
allowed lengths of strings are again  Fibonacci numbers: $Q_{k} =
F_{k}$. The considered branch of hierarchical tree connects edges
($\tau=-1,\kappa=1,\mu=1$) of the lowest bands of generations
$\eta_{3k}=F_{3k-1}/F_{3k}$.  Their string content consists of pairs of
strings with lengths $2l_n+1=F_{3n+1}$, $n=0,1,\ldots,k-1$, parities
$v_{l_n}=+1$ and inverted centers $x_{l_n}$ and $x_{l_n}^{-1}$. According
to the string hypothesis the wave function of this state is
\begin{eqnarray}
 \label{emp-wf-bottom}
    \Psi(z|\eta_{3k}) \approx \prod_{n=0}^{k-1} \prod_{j=-l_n}^{l_n}
    (z-x_{l_n}q_{l_n}^j)(z-x_{l_n}^{-1}q_{l_n}^j). 
\end{eqnarray}
Centers of the strings $x_{l_n}$ are close to 1 but can not be obtained
from the string hypothesis only. We find empirically (see
Fig.\ref{fig.3455}) that for this branch they can be fit by
$x_{l_n}\approx 1+\frac{\xi}{F_{3n+1}}$ with $\xi=1.14$. Equation 
(\ref{emp-wf-bottom}) in the limit  $k\rightarrow\infty$ gives the wave
function of the lowest energy level  of the Harper equation
with $\eta=(\sqrt{5}-1)/2$---golden mean.
More accurately, it gives an asymptotically exact value for 
$\frac{\Psi(z|\eta_{3k+3N}}{ \Psi(z|\eta_{3k}}$ 
at $k\rightarrow\infty$.

As an another example, let us consider a branch of the 
hierarchical tree for the golden mean,
which goes through levels 
$r_{2k}=\frac{2F_{2k}-F_{2k-1}}{5}+\frac{5+(-1)^k}{10}$ 
of generations $\eta_{2k}=F_{2k-1}/F_{2k}$. 
Here $k=1,2,\ldots$ and $r_{2k}$ is the number
of the level counted from the bottom of the spectrum. 
This branch  leads to the point of the spectrum characterized 
by an irrational number (filling factor)
$\zeta\equiv\lim_{k\rightarrow\infty}r_{2k}/F_{2k}
=(5-\sqrt{5})/10$. 
Each state along the branch consists of strings of lengths
$2s_n+1=F_{2n+1}$ and parities $v_{s_n}=(-1)^{3\{n/3\}}$, where 
$n=0,1,\ldots,k-1$ and  $\{\cdot\}$ denotes fractional part. 
The wave function along the branch is
\begin{eqnarray}
 \label{emp-wf-somewhere}
    \Psi(z|\eta_{2k}) \approx \prod_{n=0}^{k-1} \prod_{j=-s_n}^{s_n}
    (z-v_{s_n}q_{s_n}^j). 
\end{eqnarray}

The iterative procedure allows one to generate the string 
content and wave functions out of two irrational numbers: 
a period $\eta$ and a filling factor $\zeta$.

4. {\it Numerical evidence for the string hypothesis.} We now present a
fragment of an extensive numerical analysis that we have performed in
order to check the string hypothesis. Let us consider the bottom edge of
the lowest energy band for $\eta_9=34/55$ from the golden mean
sequence. According to (\ref{emp-wf-bottom}) the string content of this
level consists of strings of lengths $(1,1,5,5,21,21)$. All string
parities in this example are $+1$. The actual numerically calculated
roots are shown in Fig.\ref{fig.3455}. One notices immediately that
the roots organize themselves into distinct groups whose members have
roughly the same radius: two groups with 1 root($+$), two with 5
roots($\times$), and two groups with 21 roots($*$). The centers of
strings are $x_{l_n}= 2.10$, $1.23$, $0.81$, $1.05$, $\approx
1+1.14/F_{3n+1}$. The roots produced by 
eq.(\ref{emp-wf-bottom})(\fbox{$\cdot$}
on Fig.\ref{fig.3455}) are in a good agreement with the numerical
results. In fact one can prove that their accuracy of each root of the
string of the length $2l+1$ is ${\cal O}( 1/l^{2})$.

The overlap between the asymptotic ansatz wave function
(\ref{emp-wf-bottom}) and the direct numerical solution of the
eq.(\ref{HamChiral223}) for $\eta_{15}=\frac{610}{987}$ is
$\langle\psi_{\rm direct}|\psi_{\rm ansatz}\rangle\approx 0.99475$. To
see how the accuracy of ansatz (\ref{emp-wf-bottom}) is increasing with
the length of the string we replace all roots of ansatz wave function
$\psi_{\rm ansatz}$ except for the ones belonging to two largest strings
of the length $377$ by numerically obtained roots. The new wave function
$\psi_{\rm interm}$ is in some sense intermediate between pure ansatz
$\psi_{\rm ansatz}$ and numerical $\psi_{\rm direct}$ wave functions. The
overlap $\langle\psi_{\rm direct}|\psi_{\rm interm}\rangle \approx
0.999435$. This increase of an accuracy of approximation is due to a
higher accuracy of the long strings of the length $377$ with respect to
the shorter ones.

5. {\it Gaps.} A direct application of the string hypothesis is the
calculation of the gap distribution $\rho(D)$, i.e., the number of gaps
with magnitude between $D$ and $D+dD$. To find the gap distribution we
estimate the size of the smallest gap in the spectrum of Harper
equation with $\eta=P/Q$. Let us, first, consider  an arbitrary gap
of a generation with  $P$-even and  trace the genealogy of lower and
upper edges ($\tau=-1$) of the gap along two paths $J_-$ and $J_+$ of
the tree, until we find the nearest common ancestor at the generation
with denominator $Q'$. The minimal gap would correspond to the shortest
paths $J_{\pm}$, i.e., when $Q'$ is the denominator of the generation
which is parent of $P/Q$. The string decompositions of the two edges have
a common part consisting of strings of length smaller than $Q'$ (they
correspond to the common path from the origin of the tree to the parent
generation $Q'$) and {\em different strings} with lengths bigger than
$Q'$ belonging to the paths $J_{\pm}$. Then according to (\ref{en1}), the
width of the gap is the difference between energies of strings with
length greater than $Q'$ along the path $J_-$ and the energies of
strings along $J_+$. The energy of a string with a spin $l$ which
connects the center of the band ($\tau=+1$) with the edge of the band
($\tau=-1$) is
\begin{eqnarray}
    \varepsilon_l &=& 2i(q-\frac{1}{q})
    \sum_{k=-l}^{l}x_lv_lq_l^{k}
    +4(q^{1/2}+\frac{1}{q^{1/2}})
    \sim \frac{1}{Ql}
 \label{energy03} 
\end{eqnarray}
For a typical $\eta$ the length
of the largest string $2l+1$ scales as $Q$ along the sequence of rational
approximants, which gives an estimate $\varepsilon_l\sim 1/Q^2$. The
difference of energies of strings (i.e., the width of the minimal gap) is
of the same order. This gives the size of the minimal gap $D_{\rm
min}\sim 1/Q^2$. 

While moving along the tree new gaps appear, while older gaps stay
approximately the same. There is no overlap between bands---all gaps are
open (except for $\eta$ having even denominators and  $E=0$ where bands
touch)\cite{Mouche89}. Therefore the full set of gaps in the spectrum for
irrational $\eta_{\rm irr}$ consists of $Q-1$ gaps which has been opened
at the period $\eta=P/Q$ plus the smaller gaps which open at rational
approximants to $\eta_{\rm irr}$ with denominators higher than $Q$. One
can write $\int^{\infty}_{D_{\rm min}(Q)}\rho(D)dD=Q-1\sim Q$.
Assuming smooth power law gap distribution $\rho(D)\sim D^{-\gamma}$ one
obtains $\gamma=-3/2$ and the gap distribution function: 
$\rho(D)\sim D^{-3/2}$. This confirms numerical analysis of
Ref.\cite{GKP91}.

6. {\it Stating the problem: scaling hypothesis.} The string hypothesis
solves the Bethe Ansatz equations with an accuracy ${\cal O}(Q^{-2})$
for roots belonging to the largest strings. This gives asymptotically
exact results for the ``evolution'' along the hierarchical tree in the
incommensurate limit $Q\rightarrow\infty$ (see eq.(\ref{PsiAnsatz})). 
However, the most interesting quantitative characteristics of the
spectrum are actually in the finite size corrections of the order of
$Q^{-2}$ to the bare value of strings. Among them are the anomalous
dimensions of the spectrum. Let us choose a branch ${\cal J}$ of the
hierarchical tree and consider energies of the states along the branch
$E_j({\cal J})$. The scaling hypothesis states that they converge to a
point of the spectrum
$E({\cal J})$ such that $ Q^{2-\epsilon_{\cal J}}|E_j({\cal J})-E({\cal
J})|$ is bounded but does not converge to zero. The numbers
$\epsilon_{\cal J}$ are anomalous exponents. They depend on the branch
(signature of multifractality) and on $\eta$ (according to
ref.\cite{HirKohm89} for $\eta=\frac{\sqrt 5-1}{2}$ exponents
$\epsilon_J$ vary between $0.171$ and $-0.374$).  Can anomalous
dimensions be found analytically? They are determined by the finite size
corrections to the strings which, we believe, can be obtained via a more
detailed study of the BA equations. This is a technically involved but a
fascinating problem. Its solution may suggest the conformal field theory
approach, which has been proven to be effective for finding the finite
size corrections of integrable systems, without the actual solving the
Bethe Ansatz.

7. We would like to thank J. Bellissard, who suggested that the
noninteracting strings determine the gap distribution function and
pointed out the Ref. \cite{GKP91}, and Y. Hatsugai for inspiring numerics
during the initial stages of this project. We acknowledge useful
discussions with Y. Avron, G. Huber, S. Jitomirskaya, M. Kohmoto, Y.
Last, R. Seiler and A. Zabrodin. AGA was supported by MRSEC NSF Grant
DMR 9400379. PBW was supported under NSF Grant DMR 9509533.


\begin{thebibliography}{99}

\bibitem{Azbel64}
M.Ya. Azbel, Zh. Eksp. Teor. Fiz. {\bf 46}, 929 (1964)

\bibitem{Hofstadter76}
D.R. Hofstadter, Phys. Rev. B {\bf 14}, 2239 (1976).

\bibitem{AubryAndre80}
S. Aubry, G. Adnr\'{e}, Ann. Israel Phys. Soc. {\bf 3}, 133 (1980).

\bibitem{Simon82}
B. Simon, Adv. Appl. Math., {\bf 3}, 463 (1982).

\bibitem{Wilkinson84}
M. Wilkinson, Proc. R. Soc. London Ser. A {\bf 391}, 305 (1984);
M. Wilkinson, J. Phys. A {\bf 20}, 4337 (1987).

\bibitem{LastJit94}
Y. Last, Proc. XI Intern. Congress of Math. Phys. (Paris 1994);
S. Jitomirskaya, {\it ibid.}.


\bibitem{WiegmannZab94}
P.B. Wiegmann and A.Zabrodin, Nucl. Phys. {\bf B 422}, 495 (1994);
{\it idem} Phys. Rev. Lett. {\bf 72},
1890 (1994);

\bibitem{FadKash95}
L.D. Faddeev, R.M. Kashaev, Comm. Math. Phys., {\bf 169}, 181 (1995).

\bibitem{Kutz94} N. Kutz, Phys. Lett., {\bf A187}, 365 (1994).

\bibitem{HKW94}
The exact solution of (\protect\ref{BetheAnsatz}) 
corresponding to  $E=0$ as well as first numerical solutions of
BA have been obtained in:
Y. Hatsugai, M. Kohmoto, Y.S. Wu, Phys. Rev. Lett. {\bf 73}, 1134 (1994);
{\it idem} Phys. Rev. B {\bf 53}, 9697
(1996).

\bibitem{TakSuz72}
M. Takahashi, M. Suzuki,
Progr. Theor. Phys. {\bf 48}, 2187 (1972).

\bibitem{Mouche89}
P. v. Mouche, Commun. Math. Phys. {\bf 122}, 23 (1989).

\bibitem{ThoulessNiu83} 
D. J. Thouless, Q. Niu, J. Phys. A: Math. Gen., {\bf 16}, 1911 (1983).

\bibitem{comment}
For details about rational points see
\cite{WiegmannZab94,FadKash95,ATW97}.

\bibitem{ATW97}
A.G. Abanov, J.C. Talstra, P.B. Wiegmann, cond-mat/9711274.


\bibitem{TKNN82}
D.J. Thouless, M. Kohmoto, M.P. Nightingale, M. den Nijs,
Phys. Rev. Lett. {\bf 49}, 405 (1982).

\bibitem{DAZ85}
I. Dana, Y. Avron, J. Zak, J. Phys. C
{\bf 18}, L679 (1985).

\bibitem{Streda82}
P. Streda, J. Phys. C: Solid State Phys., {\bf 15}, L717 (1982).

\bibitem{Venkov70} B.A. Venkov,
{\it Elementary number theory}, WN, 1970.

\bibitem{MezNep90} L. Mezincescu, R. I. Nepomechie,
 Phys. Lett., {\bf B246}, 412 (1990).

\bibitem{GKP91}
T. Geisel, R. Ketzmerick, G. Petschel,
 Phys. Rev. Lett. {\bf 66}, 1651 (1991).

\bibitem{HirKohm89}
H. Hiramoto, M. Kohmoto, Phys. Rev. B{\bf 40}, 8225 (1989).

\end{thebibliography}
\end{document}